\documentclass[12pt]{article}

\usepackage{times,color}
\usepackage{amsmath}
\usepackage{subfigure,epsfig,graphicx,epstopdf}
\usepackage[colorlinks]{hyperref}
\hypersetup{linkcolor = {blue},	citecolor = {blue},	urlcolor = {blue}}
\usepackage{cite}

\newenvironment{sciabstract}{%
\begin{quote} \bf}
{\end{quote}}

\newcounter{lastnote}

\title{Observation of Magic Angle and Wall State in Twisted Bilayer Photonic Graphene}

\author{Yao Wang,$^{1,2}$ Yi-Jun Chang,$^{1,2}$ Jun Gao,$^{1,2}$ Yong-Heng Lu,$^{1,2}$\\
	 Zhi-Qiang Jiao,$^{1,2}$ Fang-Wei Ye,$^{1}$ Xian-Min Jin$^{1,2,\ast}$\\
\normalsize{$^1$School of Physics and Astronomy, Shanghai Jiao Tong University, Shanghai 200240, China}\\
\normalsize{$^2$Synergetic Innovation Center of Quantum Information and Quantum Physics,}\\
\normalsize{University of Science and Technology of China, Hefei, Anhui 230026, China}\\
\normalsize{$^\ast$E-mail: xianmin.jin@sjtu.edu.cn}\\
}

\date{}

\begin{document}
\baselineskip24pt

\maketitle

\begin{sciabstract}
Graphene, a one-layer honeycomb lattice of carbon atoms, exhibits unconventional phenomena and attracts much interest since its discovery~\cite{graphene_1, graphene_2, graphene_3, graphene_4, graphene_5, graphene_6, graphene_7, graphene_8}. Recently, an unexpected Mott-like insulator state induced by moir{\'e} pattern and a superconducting state are observed in magic-angle-twisted bilayer graphene~\cite{exp_bi_2,exp_bi_3}, especially, without correlations between electrons, which gives more hints for the understanding and investigation of strongly correlated phenomena. The photon as boson, behaving differently with fermion, can also retrieve the unconventional phenomena of graphene, such as the bearded edge state which is even never been observed in graphene due to the unstability~\cite{p_graphene_4}. Here, we present a direct observation of magic angle and wall state in twisted bilayer photonic graphene. We successfully observe the strong localization and rapid diffusion of photon at the regions with AA and AB stacking order around the magic angle, respectively. Most importantly, we find a wall state showing the photon distribution distinctly separate at the regions with AA and AB/BA stacking order in the lowest-energy band. The mechanism underlying the wall states may help to understand the existence of both Mott-like insulating state and superconducting state in magic-angle twisted bilayer graphene. The accessibility of magic angle in twisted bilayer photonic graphene adds the boson behavior into graphene superlattice and the observation of wall state will also deep the understanding of matter.\\
\end{sciabstract}


\noindent The photonic graphene, constructed using evanescently coupled waveguide array, makes it convenient and easily accessible to experimentally investigate the graphene physics~\cite{p_graphene_1, non_graphene_1, non_graphene_2, p_graphene_2, p_graphene_3, p_graphene_4, p_graphene_5}. Besides the physically scalable and structurally designable of waveguide system~\cite{QW_2D,FH,PIT_Gap}, photonic graphene also allow one to directly observe the wave and particle dynamics of electrons in graphene, due to the mathematically equivalence of paraxial wave equation describing propagation of photon in waveguide array and the Schr{\"o}dinger equation describing time evolution of electrons~\cite{p_graphene_4}. Based on the photonic waveguide platform, photonic graphene provides an elegant way to investigate the features in graphene, such as nonlinear domain~\cite{non_graphene_1, non_graphene_2}, multi-particle regime~\cite{Qutop_2photon}, and direct observation of the actual wavefunction~\cite{p_graphene_4, wave_1,wave_2}, which are virtually impossible in electronic systems.

Beyond the one-layer graphene, more novel and interesting features are predicted~\cite{th_bi_1, th_bi_2, th_bi_3, th_bi_4, th_bi_5} and demonstrated~\cite{exp_bi_1, exp_bi_2, exp_bi_3, exp_bi_4, exp_bi_5, exp_bi_6} in the twisted bilayer photonic graphene, such as unconventional superconductivity~\cite{exp_bi_2} and correlated insulator behavior~\cite{exp_bi_3} in magic-angle graphene superlattices, paving an accessible way to study strongly correlated phenomena and other exotic many-body quantum phases~\cite{correlation_1, correlation_2, correlation_3, correlation_4}. Similar to the realization of photonic graphene, the promotion of phenomena in magic-angle-twisted bilayer graphene from electronic systems to boson systems will facilitate the investigation of the emerging field in a fully controllable fashion.

Here, we experimentally observe the magic angle and wall states in twisted bilayer photonic graphene. Using the femtosecond laser direct writing technique, we are able to map the designed magic-angle-twisted bilayer photonic graphene into borosilicate silica. The injected photons at the regions with AA stacking order are strongly localized showing the Mott-like insulating state, while the photons at the regions with AB stacking order are rapidly diffuse showing the conducting state. More interestingly, we find a set of specific states in the lowest-energy band, namely the ``wall states'', separating above two regions. The observation of magic angle in twisted bilayer photonic graphene promotes the bilayer graphene from fermion systems to boson systems, and the demonstrated wall states will deep the understanding of this field.


In the photonic waveguide array, the evolution of photon is governed by
\begin{equation}
	i\partial_z\psi_{n}(z) = \sum_{m}H_{nm}^{\text{TB}}\psi_{m}(z),
	\label{Eq_evo}
\end{equation}
where $\psi_{n}(z)$ is the wavefunction of $n$th waveguide mode as the function of evolution distance $z$, and $H_{nm}^{\text{TB}}$ is the discrete system Hamiltonian obtained from continuum Hamiltonian in the paraxial wave equation by applying the tight-binding approximation, as presented in Supplementary Materials. Due to the fact that the coupling exist only between nearest-neighbor waveguides, there are no higher order items of potential well and coupling in our system, which well matches the tight-binding model and gives an accurate analogy for the bilayer graphene in electronic systems.

\begin{figure}[htbp]
	\centering
	\includegraphics[width=0.6\linewidth]{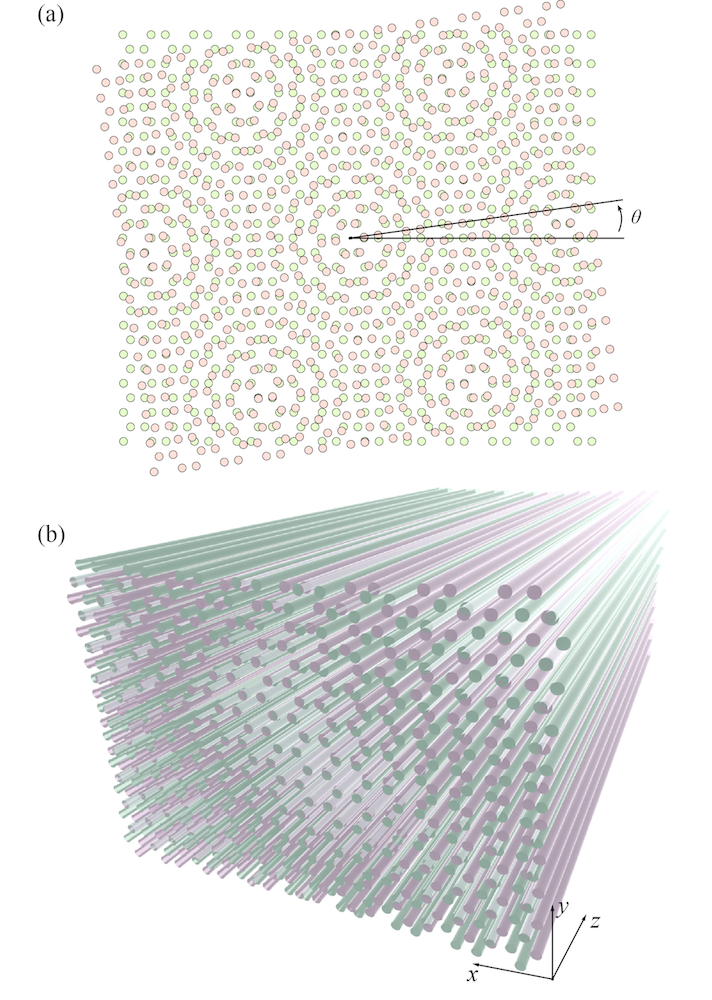}
	\caption{\textbf{Schematic of twisted bilayer photonic graphene.} \textbf{(a)} The bilayer graphene superlattices with a twisted angle of $\theta$. The sites in different layer (marked in green and pink respectively) are staggered and the moir{\'e} pattern appears clearly. \textbf{(b)} Schematic of the fabricated lattice on a photonic chip. We design and fabricate the twisted bilayer photonic graphene by mapping the superlattice structure to a two-dimension photonic lattice. The sites belonging to different layer are marked in green and pink respectively.}
	\label{f1}
\end{figure}

As sketched in Fig.\ref{f1}(a), in twisted bilayer graphene with an angle $\theta$,  the corresponding sites in different layers are staggered and the moir{\'e} pattern appear clearly. Such that, we are able to mapping the final structure into a two-dimension lattice and construct it using waveguides as illustrated in Fig.\ref{f1}(b), where the $z$ direction corresponds to the evolution time, and the waveguides in different layers are presented with green and pink respectively. We implement the twisted bilayer photonic graphene on a photonic chip by using femtosecond laser direct writing technique~\cite{fabri_1,fabri_2,fabri_3,fabri_4}, and the nearest-neighbor spacing is designed as 15 $\mu$m corresponding to a coupling strength of 0.19 mm$^{-1}$ for the photon at wavelength of 810 nm. The array contains more than 600 waveguides and the evolution distance is set as 30 mm. More details about the waveguide and lattice are shown in Supplementary Materials.

\begin{figure}[htbp]
	\centering
	\includegraphics[width=0.65\linewidth]{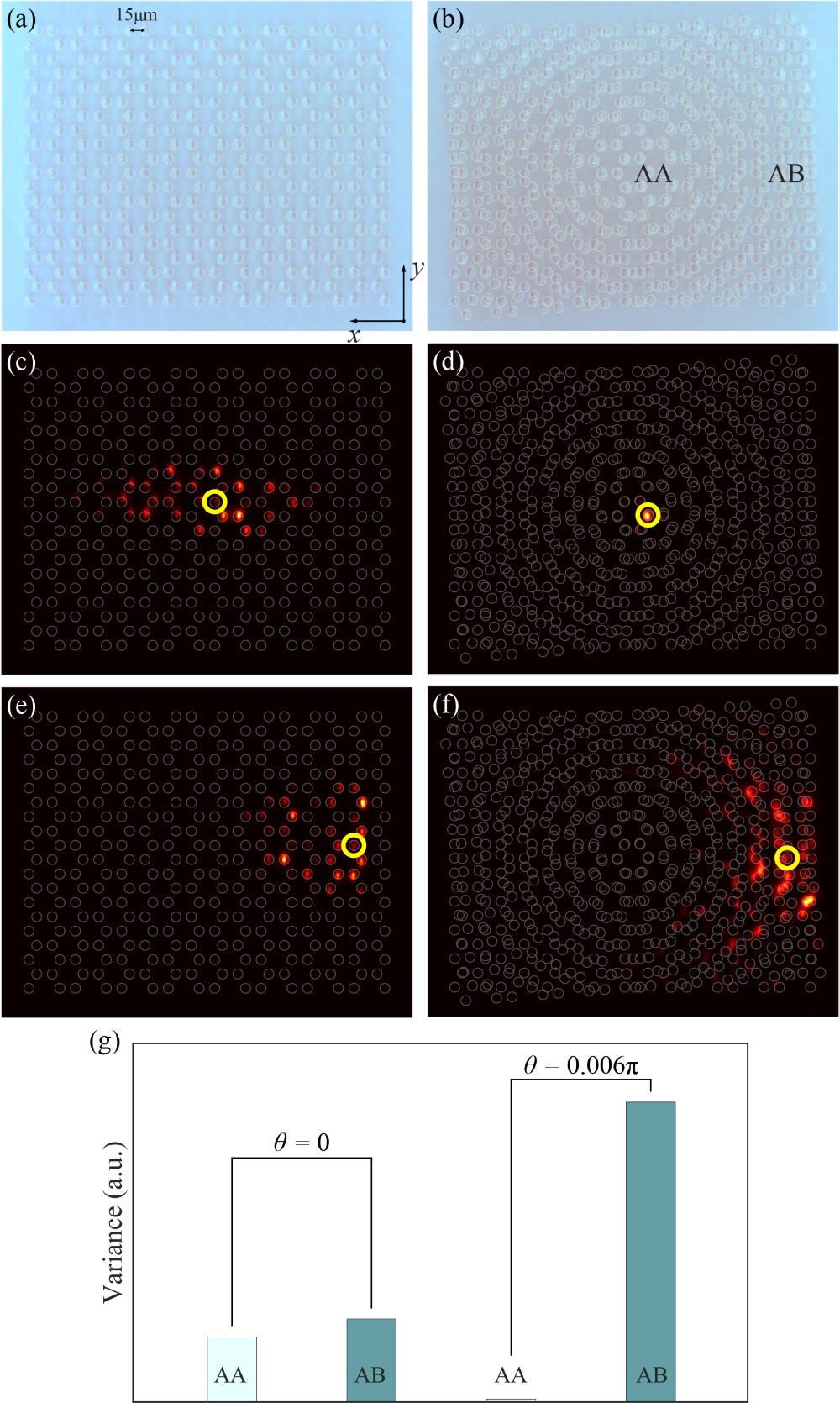}
	\caption{\textbf{Measured wavefunction obtained from the output facet.} Microscope images of the output facet of femtosecond-laser-written bilayer photonic graphene with twisted angle of 0 \textbf{(a)} and 0.006$\pi$ \textbf{(b)}. The measured wavefunction for the photon injected at the regions of AA stacking order \textbf{(c)} and AB stacking order \textbf{(e)}. The wavefunction result of magic-angle-twisted bilayer photonic graphene for the photons injected at the regions of AA stacking order \textbf{(d)} and AB stacking order \textbf{(f)}. The data of the photon wavefunction is normalized by the highest value. \textbf{(g)} The variance of the photon distribution, where $\sigma^{0.006\pi}_\text{AB} \gg \sigma^0_\text{AA} \approx \sigma^0_\text{AB} \gg \sigma^{0.006\pi}_\text{AA} \approx 0$. }
	\label{f2}
\end{figure}

We show the microscope images of the output facet of the implemented twisted bilayer photonic graphene with the effecitve angle $\theta=0$ and $0.006\pi$ in Figs.\ref{f2}(a-b) respectively. As theoretically predicted and experimentally demonstrated in electronic system~\cite{th_bi_2, exp_bi_2, exp_bi_3}, the first magic angle appears at twist angles of about $0.006\pi$, and the regions with AA and AB stacking order could be clearly identified as marked in Fig.\ref{f2}(b). We inject photons into the twisted bilayer photonic graphene from sites at the regions with AA and AB stacking order respectively, which is marked with yellow circles. As the results shown in Figs.\ref{f2}(d, f), for magic-angle-twisted bilayer photonic graphene, the photons are strongly localized at the region with AA stacking but rapidly diffuse at the region with AB stacking. As a comparison, the photons in zero-angle-twisted bilayer photonic graphene are steadily diffuse either at the regions with AA or AB stacking order, as shown in Figs.\ref{f2}(c, e).

To quantify the diffusion of photons, we define a value $\sigma^\theta_\text{order}$ to describe the variance of the distribution of photons as $\sigma^\theta_\text{order}=\sum_{i}\varDelta l_i^2I_{i}$, where $\varDelta l_i$ is the normalized spacing between the site $i$ and the original excited site, $I_m$ is the intensity of the photons in the site $i$, and the order means the stacking order. The results are shown in Fig.\ref{f2}(g), as we observed from the evolution distribution, the variances of photon distribution at the regions with AA and AB stacking order are almost equal for zero-angle-twisted bilayer photonic graphene with $\sigma^0_\text{AA}\approx\sigma^0_\text{AB}$. For magic-angle-twisted bilayer photonic graphene, especially, the result of $\sigma^{0.006\pi}_\text{AB}\gg\sigma^0_\text{AB}$ demonstrates the unconventional conductivity observed in electronic system, and $\sigma^{0.006\pi}_\text{AA}\approx0$ exhibits the localization of the photons corresponding to the correlated insulator behavior.

\begin{figure}[htbp]
	\centering
	\includegraphics[width=1.0\linewidth]{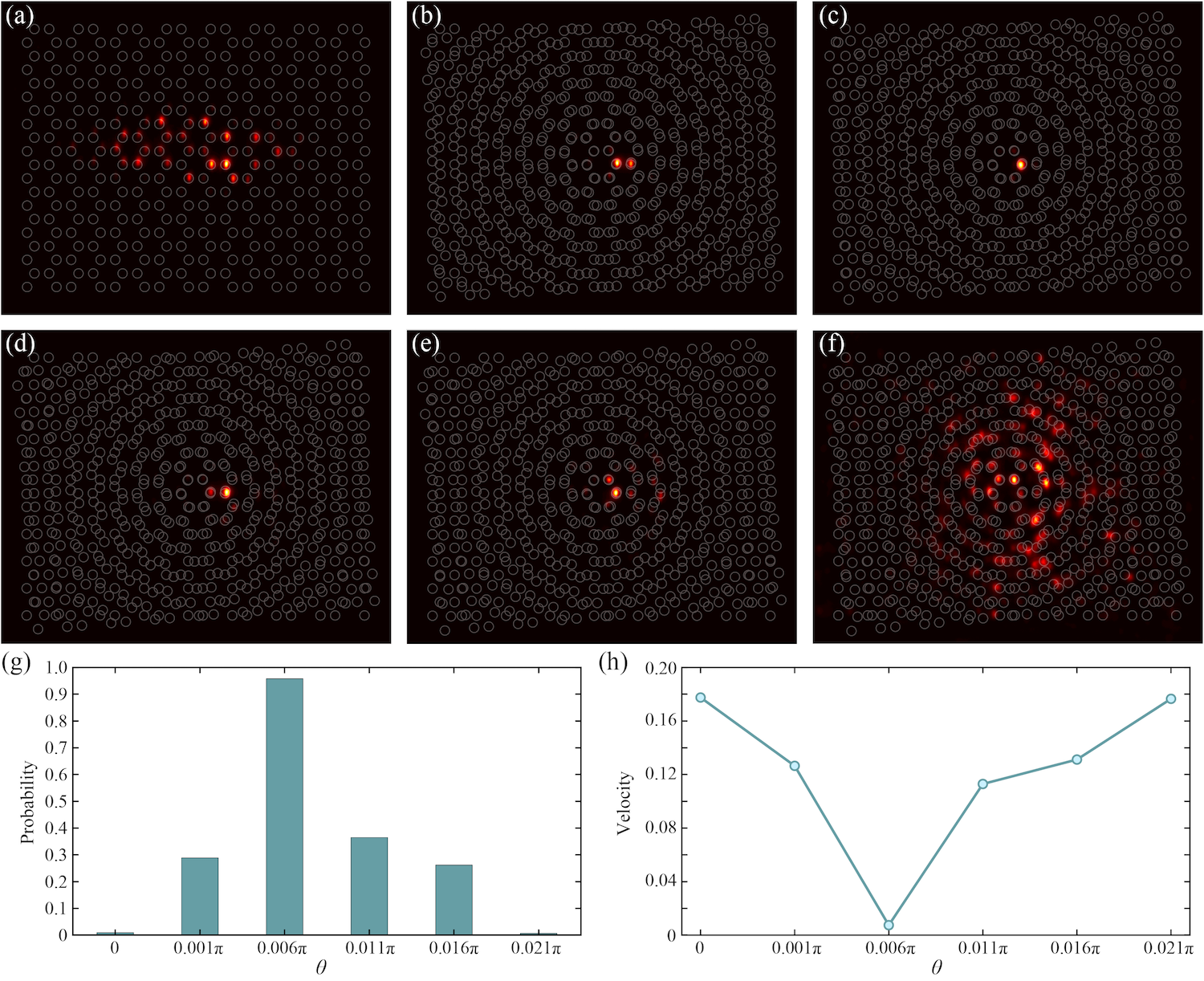}
	\caption{\textbf{The identification of the magic angle.} \textbf{(a-f)} The measured wavefunction obtained from the output facet for the bilayer photonic graphene with twisted angle of 0, 0.001$\pi$, 0.006$\pi$, 0.011$\pi$, 0.016$\pi$, and 0.021$\pi$. \textbf{(g)} The generalized return probability obtained from the wavefunction. The value of the return probability for the magic-angle-twisted bilayer photonic graphene approaches to 1. \textbf{(h)} The photon diffusion velocity. The value goes down to zero and then up when the twisted angle goes pass through magic angle $\theta=0.006\pi$.}
	\label{f3}
\end{figure}

To further determine that the above phenomena appear only in the magic-twisted angle, we fabricate different twisted bilayer photonic graphene arrays with effective angles ranging from 0.001$\pi$ to 0.021$\pi$ with a step of 0.005$\pi$. As the results shown in Figs.\ref{f3}(a-f), we inject the photons into the twisted bilayer photonic graphene from the site at the region with AA stacking order, only the magic-angle-twisted bilayer photonic graphene prevents the photons from diffusing. To quantify the fraction of photon probability remaining confined on the injecting site, we calculate the generalized return probability $\xi_j^\theta$ defined by $\xi_j^\theta=\sum_{j-d}^{j+d}I_i/\sum_{1}^{N}I_i$, which quantifies the probability of the photons that remain within a small distance $d$ from the injection site $j$ among all $N$ sites, and $d$ is adopted as 0 here. We show the calculated results in Fig.\ref{f3}(g), there are almost no photon left in the injecting sites for the twisted bilayer photonic graphene with angle of 0 and 0.024$\pi$, while the result of $\xi_j^{0.006\pi}\approx1.0$ verifies the results obtained from observed wavefunction emerging from the output facts.

As the theoretical prediction, the renormalized Fermi velocity pass through zero at the magic angle, which means the electrons would not diffuse to elsewhere anymore, similar with the observed photon evolution in our experiment. We also could quantify the photon diffusion velocity in twisted bilayer photonic graphene, with the direct way, as $v^\theta_p=(1-\xi_j^\theta)/z\bar{c}$, where $z\bar{c}$ is the normalized propagation distance and approximate to 5.6 in our experiment. We show the measured photon diffusion velocity in Fig.\ref{f3}(h), as expected, $v^\theta_p$ goes down to zero and then up when the twisted angle goes pass through magic angle $\theta=0.006\pi$.

\begin{figure}[htbp]
	\centering
	\includegraphics[width=1.0\linewidth]{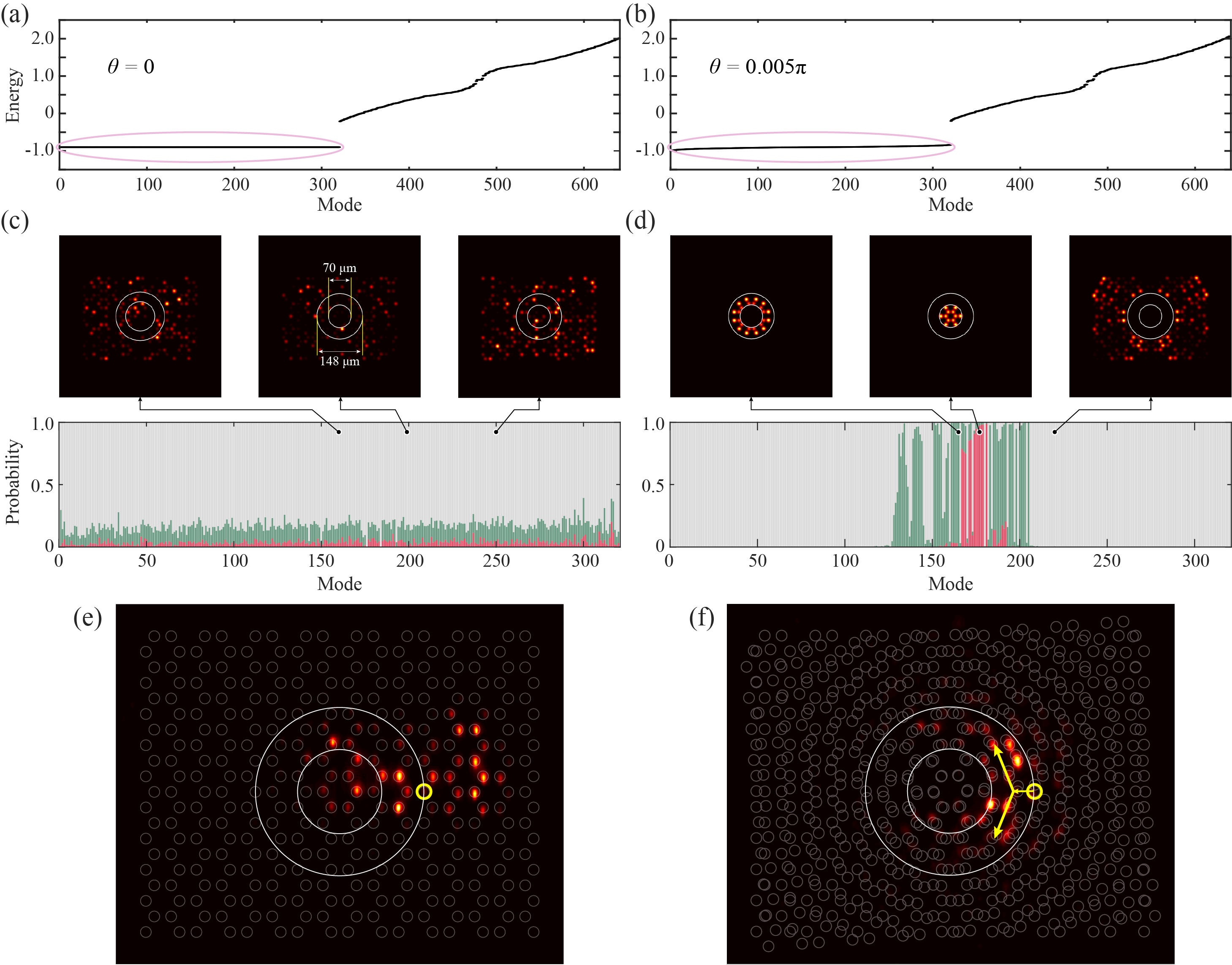}
	\caption{\textbf{Experimental observation of the wall state.} \textbf{(a-b)} The spectrum of the finite-size bilayer photonic graphene with twisted angle of 0 and 0.005$\pi$. The pink ellipses mark the lowest-energy band. \textbf{(c-d)} Spatial distribution for the eigenmode in the lowest-energy band. The photons occupy all three areas simultaneously for the twisted angle of 0 (c), but occupy three regions independently for the twisted angle near magic angle (d). The insulating area is the scope with a distance smaller than 35 $\mu$m from the lattice center (red bar), the conducting area is the scope with a distance larger than 74 $\mu$m from the lattice center (gray bar), and the scope between them belongs to wall (green bar). \textbf{(e-f)} The observed wall state. The photons are injected into the lattice from the site in the outer boundary of the wall, and spread to all three areas when the twisted angle is 0 but almost only diffuse into the wall when the twisted angle goes near to magic angle.}
	\label{f4}
\end{figure}

As the special capability of twisted bilayer photonic graphene, it allows us to direct observe the wave and particle dynamics of photons. We further calculate the spectrum of the implemented finite-size graphene superlattice to find more underlying physics, which is shown in Figs.\ref{f4}(a-b). The gap keeps open and the spectrum seems similar when we tune the twisted angle. However, the evolution of lowest-energy band structure of twisted bilayer graphene near the magic angle in momentum space reported in Ref.~\cite{exp_bi_3} inspires us to reveal the hidden mechanisim in the lowest-energy band of the spectrum.

The spatial distribution of the eigenmode gives the photon distribution of each basis state defined as $D_i(E)=\sum_{m}\delta(E-E_m)|\psi_m^{(i)}|^2$, where $E_m$ is the energy of the $m$th eigenmode $\psi_m^{(i)}$ and $i$ is the site label. We find a special wall state in the twisted bilayer photonic graphene near the magic angle. The wall states separate the superlattice into three areas in our finite-size lattice, namely insulating, wall, and conducting area. We calculate the photon distribution percentage of the three areas for each eigenmode in the lowest-energy band, and the results of twisted angle with 0 and 0.005$\pi$ are presented in Fig.\ref{f4}(c) and Fig.\ref{f4}(d) respectively. The photons spread into all three areas in all the mode for the twisted angle with 0, while the photons almost distribute in only one of the areas for the twisted angle with 0.005$\pi$. The spatial distributions of all mode in the lowest-energy band are presented in Supplementary Materials. The wall states in magic-angle-twisted bilayer photonic graphene may help to understand why both the Mott-like insulator state and conducting state simultaneously exist in the superlattice.

In our experiment, we inject the photon from the outer boundary of the wall, and observe the wavefunction of the photon after an evolution of 30 mm. As shown in Figs.\ref{f4}(e-f), the photons diffuse to three areas for the twisted angle with 0 but almost only occupy in the wall for the magic-angle-twisted bilayer photonic graphene. From the wavefunction emerging in the output facet, we can find that the photons diffuse into the inside region of the wall and then evolve along the wall range in the magic-angle-twisted bilayer photonic graphene, while the photons diffuse into all directions for the zero-twisted angle. The results confirm the existence of the wall states in experiment, which may provide a better understanding to the unconventional phenomena in magic-angle graphene superlattices.

The observation of insulating, wall, and conducting area may can be derived from the result of the local density of states and implies the electron behavior in twisted bilayer graphene. It should be noticed that the sites at the region with AB/BA stacking order are connected all over the lattice, while the sites at the region with AA stacking order are isolated from each other. The electrons can freely move at the regions with AB/BA stacking order, which leads to the superconducting state in magic-angle twisted bilayer graphene. The strongly concentrated electron density at the region with AA stacking order render to the Mott-like insulating state, and the wall states observed in our experiment separate it from the conducting state. They exist at the same time but be dominant respectively under different environments including the temperature and carrier density.



In conclusion, we present an experimental observation of magic angle and wall state in the twisted bilayer photonic graphene. Both the insulator state and the conducting state are observed in our bosonic system. The photons at the region with AA stacking order are strongly localized and have diffusion velocity of nearly zero, and the photons have a particularly high diffusion velocity and spread rapidly at the region with AB stacking order. Importantly, we find a wall state that has not been observed before, the wall states separate the superlattice to three areas and may help to understand the existence of both Mott-like insulator state and the superconducting stat in magic-angle-twisted bilayer graphene.

Based on the photonic waveguide system, we are able to directly observe the wavefunction of the system in controllable fashion, which provides an alternative but accessible approach for investigating the twisted bilayer graphene physics. The observation here also enriches and promotes this field into bosonic system, which, together with the newly found wall state, will deep the understanding of matter. With the structurally designable of waveguide system, we may able to achieve the direct observation of other magic angles~\cite{th_bi_2} in twisted bilayer graphene in near future. Besides, the newly found wall state may also exist in other lattice structure~\cite{Pythagorean}, and may promote applications in robust manipulation of structured light field, for example, supporting and preserving orbital angular momentum of light on a photonic chip~\cite{OAM}.\\

\subsection*{Acknowledgments}
The authors thank Roberto Osellame and Jian-Wei Pan for helpful discussions. This research is supported by National Key R\&D Program of China (2017YFA0303700), National Natural Science Foundation of China (NSFC) (61734005, 11761141014, 11690033), Science and Technology Commission of Shanghai Municipality (STCSM) (15QA1402200, 16JC1400405, 17JC1400403), Shanghai Municipal Education Commission (SMEC)(16SG09, 2017-01-07-00-02-E00049). X.-M.J. acknowledges support from the National Young 1000 Talents Plan.

\subsection*{Author contributions} 
X.M.J. conceived and supervised the project.
Y.W. and Q.Z.J. fabricated the samples.
Y.W., Y.J.C., J.G., Y.H.L., Q.Z.J. and X.M.J. performed the measurements. 
Y.W. and F.W.Y conducted the theoretical analysis. 
Y.W. and X.M.J. analyzed the data and wrote the paper with input from all authors. 

\subsection*{Competing interests}
The authors declare no competing interests.

\subsection*{Data availability}
The data that support the findings of this study are available from the corresponding authors on reasonable request.

\baselineskip21pt
\clearpage

\clearpage

\clearpage

\clearpage

\clearpage

\section*{Supplementary Materials: Observation of Magic Angle and Wall State in Twisted Bilayer Photonic Graphene}
	
\noindent Yao Wang,$^{1,2}$ Yi-Jun Chang,$^{1,2}$ Jun Gao,$^{1,2}$ Yong-Heng Lu,$^{1,2}$ Zhi-Qiang Jiao,$^{1,2}$ Fang-Wei Ye,$^{1}$ Xian-Min Jin$^{1,2,\ast}$\\
$^1$School of Physics and Astronomy, Shanghai Jiao Tong University, Shanghai 200240, China\\
$^2$Synergetic Innovation Center of Quantum Information and Quantum Physics,\\
${\ \ }$University of Science and Technology of China, Hefei, Anhui 230026, China\\
$^\ast$E-mail: xianmin.jin@sjtu.edu.cn\\

\baselineskip24pt

\subsection{The derivation of evolution equation for photonic lattice}
The equation in the main text describing the evolution of photon in our system is derived from the Helmholtz equation. In this section we give the derivation of general case.	Considering the waveguide (with electric permittivity $\varepsilon$ and magnetic permeability $\mu$) in the substrate (with electric permittivity $\varepsilon_0$ and magnetic permeability $\mu_0$), the Helmholtz equation can be written as
\begin{equation}
	\nabla^{2} E+k_{0}^{2} n^{2}(x, y, z) E=0 \quad (k_{0}^{2}=\omega^{2} \mu_{0} \varepsilon_{0} ,\quad n^{2}=\frac{\mu \varepsilon}{\mu_{0} \varepsilon_{0}})
	\label{Hel}
\end{equation}
By inserting the solutions $E(x, y, z)=\psi(x, y, z) e^{i(\beta z-\omega \tau)}$, where $\beta=n_{0} k_{0}$ and $\omega=c k_{0} / n_{0}$ into Eq.\ref{Hel} and ignoring the quadratic term of $z$, we can obtain 
\begin{align}
	\label{continue}
	i \partial_z\psi(x, y, z)&=-\frac{1}{2 k_{0}}\nabla^{2} \psi(x, y, z)-\frac{k_{0} \Delta n(x, y, z)}{n_{0}} \psi(x, y, z)\\
	&=H\psi(x, y, z).
\end{align}
The obtained paraxial wave equation Eq.\ref{continue} is similar to the Schr{\"o}dinger equation as
\begin{equation}
	i\hbar\partial_t\psi = -\frac{\hbar^2}{2\mu}\nabla^{2}\psi - U(x, y, z)\psi(x, y, z),
	\label{Sch}
\end{equation}
which gives the mathematically equivalence.	To describe the evolution of photons in waveguide system, we can employ the tight-binding approximation to the result for the waveguide system, and we obtain
\begin{align}
	i\partial_z\psi_n&=-\frac{1}{2 k_{0}}\left(\frac{\psi_{n-1}-2 \psi_{n}+\psi_{n+1}}{d^{2}}\right)-\frac{k_{0} \Delta n(x, z)}{n_{0}} \psi_{n}\\
	&=-\frac{1}{2 k_{0} d^{2}}\left(\psi_{n-1}+\psi_{n+1}\right)-\left(\frac{k_{0} \Delta n(x, z)}{n_{0}}-\frac{1}{k_{0} d^{2}}\right) \psi_{n}\\
	&=-c\left(\psi_{n-1}+\psi_{n+1}\right)-\beta \psi_{n}\\
	&=H^\text{TB}\psi_{n},
\end{align}
where the $c$ is the coupling coefficient between the adjacent sites and $\beta$ is the on-site energy. Thus, we finish the derivation of evolution equation from Helmholtz equation.
	
\subsection{The details of the waveguide and the lattice}
\begin{figure}
	\centering
	\includegraphics[width=0.8\linewidth]{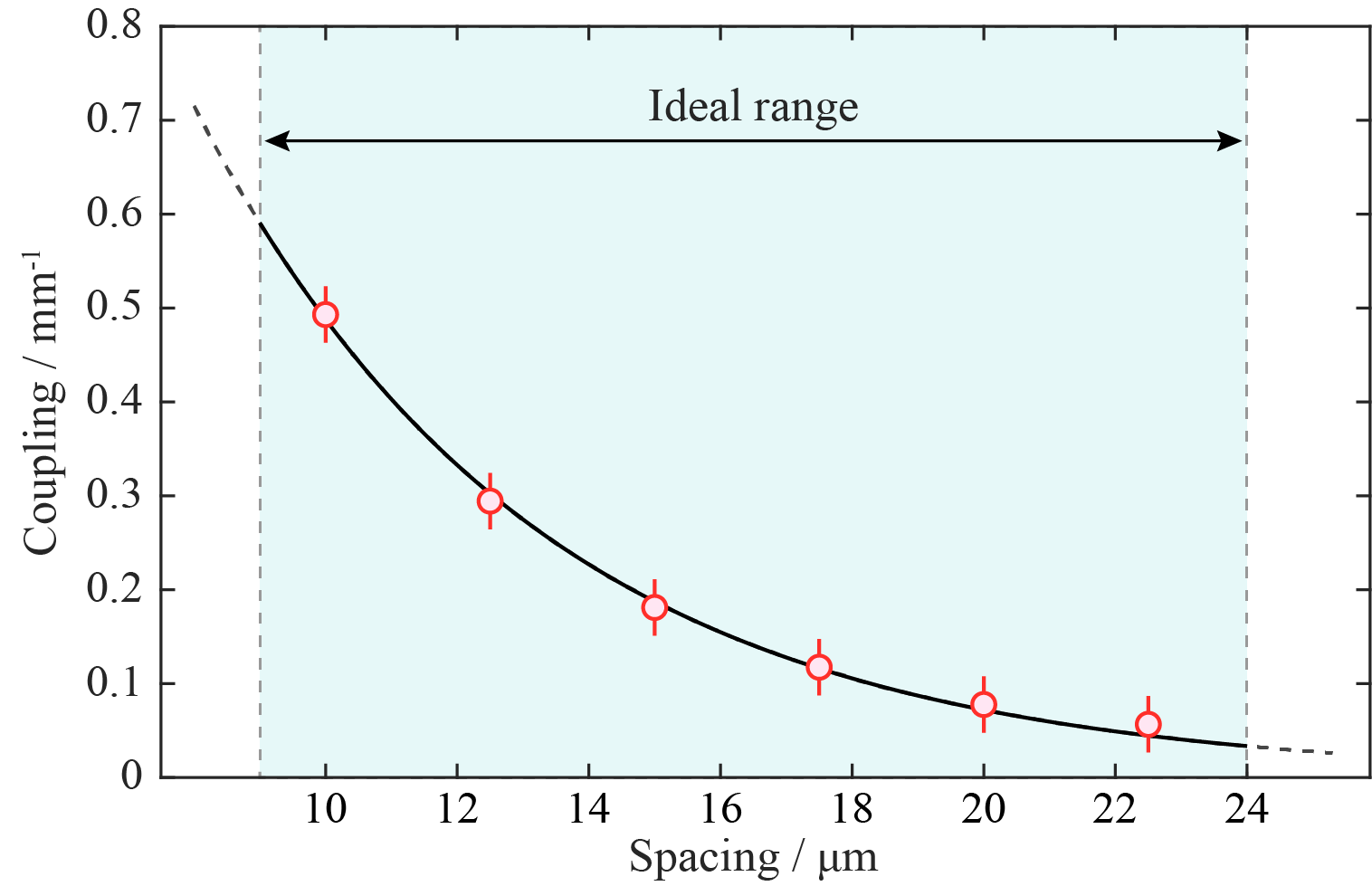}
	\caption{\textbf{The relationship between the spacing and coupling.} The ideal spacing range for the evanescent wave coupling of photons is 9--24 $\mu$m, which is marked in green. The red dots are the measured values, and the black line is the fitting result.}
	\label{s_coupling}
\end{figure}

In this section, we will give more detailed description on the waveguide and the lattice. Using the femtosecond direct laser writing technique, we are able to fabricate the designed arbitrary lattices. In our experiment, the laser-written single waveguide are elliptical with the horizontal and vertical diameters of 3 and 10 $\mu$m, respectively. The refractive index functional form of the single waveguide can be described as $\Delta n(x ,y) = n_0e^{-((2x/d_H)^2+(2y/d_V)^2)^3}$, and the anisotropy in the inter-waveguide coupling induced by the ellipticity is small and can be ignored~\cite{waveguide}. In our system, the spacing range achieving ideal evanescent wave coupling of photons between the adjacent sites is 9--24 $\mu$m, as shown in Fig.\ref{s_coupling}.

\begin{figure}
	\centering
	\includegraphics[width=0.95\linewidth]{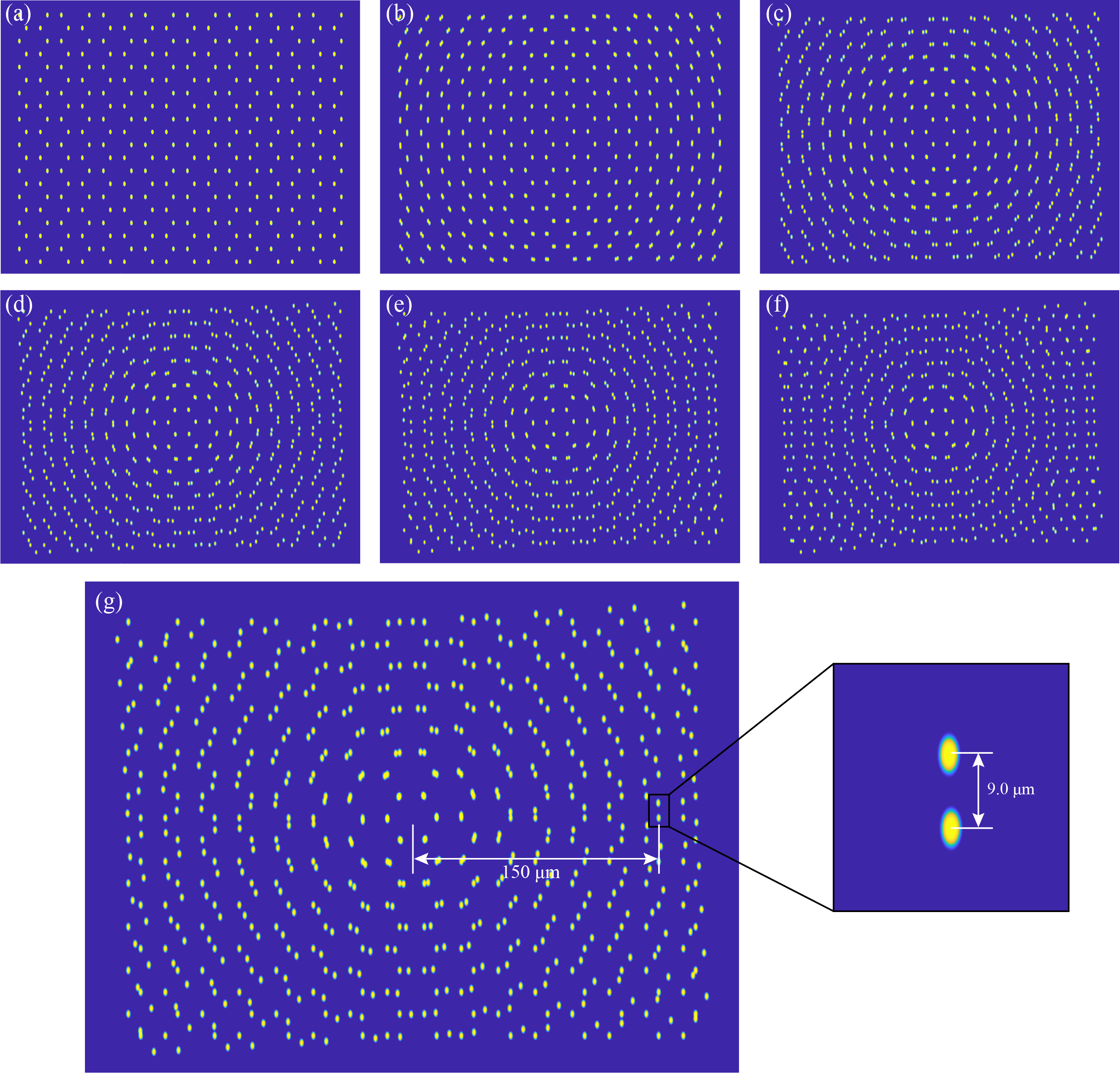}
	\caption{\textbf{The refractive index distribution of bilayer photonic graphene lattice with different twisted angles.} \textbf{(a-f)} The refractive index distribution of lattices. The twisted angle is adopted ranging from 0 to 0.025$\pi$ with a step of 0.005$\pi$ for the results in (a)-(f). \textbf{(g)} The refractive index distribution of bilayer photonic graphene lattice with twisted angle of 0.019$\pi$. The staggered distance between the corresponding sites 150 $\mu$m away from the twisted center in different layers is 9.0$\mu$m.}
	\label{s_angle}
\end{figure}

As we mentioned in the main text, the twisted bilayer photonic graphene is obtained by mapping the two-layer lattice to a two-dimension lattice. Though the sites in different layer have staggered, there still some sites overlap with each other during the mapping process, especially, when the twisted angle is small. We show the refractive index distribution in Figs.\ref{s_angle}(a-f), in which we can find that almost all the sites are overlapped for small twisted angle [Figs.\ref{s_angle}(a-c)]. In other words, there is a angle $\theta_0$ that the lattice is actually the single-layer photonic graphene when the twisted angle smaller than it. The effective twisted angle for the bilayer photonic graphene should be obtained as $\theta = \theta_d-\theta_0$, where $\theta_d$ is the designed twisted angle.

The angle $\theta_0$ is a relative value, and also is related to the lattice size and the waveguide properties. The physics of twisted bilayer photonic graphene is not influenced by the specific value of $\theta_0$. In our experiment, to identify the $\theta_0$, we investigate the staggered distance of the sites 150 $\mu$m away from the twisted center, which are the sites in the region with AB stacking order and are adopted as the exciting site mostly near to the edge of lattice. As mentioned above, the minimum spacing for the ideal evanescent wave coupling is 9 $\mu$m, hence we adopt the staggered distance of 9 $\mu$m as the critical condition of the $\theta_0$. As shown in Figs.\ref{s_angle}(g), we obtain the value of $\theta_0$ as 0.019$\pi$.

As mentioned above, some information is missing or modified during the mapping process, such as the loss of coupling between some sites due to the overlap and the modification of twisted angle. It may implies that this information may not the crucial point for understanding the physics behind the twisted bilayer graphene. On the other hand, the difference also is calling for more theoretical exploration.

\end{document}